\documentclass[final,1p,times]{elsarticle}
\usepackage{setspace}
\usepackage{graphics,amssymb,amsfonts, multirow,booktabs,amsmath,hyperref,epsfig,color,eurosym, amstext,color, dcolumn,adjustbox,lscape, placeins, graphicx}
\graphicspath{{figure/}}
\journal{arXiv}

\usepackage{makecell}
\usepackage[singlelinecheck=false]{caption}
\DeclareCaptionFont{tenpt}{\fontsize{10pt}{11pt}\selectfont #1}
\usepackage{float}
\hypersetup{pdfauthor={Name}}
\usepackage{subcaption}
\usepackage{endnotes}
\biboptions{authoryear}
\setcitestyle{aysep={,},yysep={;}}
\usepackage{pdflscape}
\usepackage{placeins}
\usepackage{graphicx}

\begin{document}
\begin{frontmatter}
\title{
Population Concentration in High-Complexity Regions within City during the heat wave}

\author[a]{Hyoji Choi}
\author[a]{Jonghyun Kim}
\author[b,c,a]{Donghyeon Yu }
\author[d,c,a]{Bogang Jun \corref{cor1}}
\ead{bogang.jun@inha.ac.kr}

\cortext[cor1]{Corresponding author}

\address[a]{Research Center for Small Businesses Ecosystem, Inha University, Incheon, South Korea}
\address[b]{Department of Statistics, Inha University, Incheon, South Korea}
\address[c]{Department of Data Science, Inha University, Incheon, South Korea}
\address[d]{Department of Economics, Inha University, Incheon, South Korea}

\begin{abstract} 
This study investigates the impact of the 2018 summer heat wave on urban mobility in Seoul and the role of economic complexity in the region's resilience. Findings from subway and mobile phone data indicate a significant decrease in the floating population during extreme heat wave, underscoring the thermal vulnerability of urban areas. However, urban regions with higher complexity demonstrate resilience, attracting more visitors despite high temperatures. Our results suggest the centrality of economic complexity in urban resilience against climate-induced stressors. Additionally, it implies that high-complexity small businesses' clusters can serve as focal points for sustaining urban vitality in the face of thermal shocks within city. In the long run perspective, our results imply the possibility that people are more concentrated in high complexity region in the era of global warming.
\end{abstract}

\end{frontmatter}

\newpage
\section{Introduction}
Resilience has recently gained attention in the context of understanding the development and sustainability of regional economy during the ongoing multiple crises ~\citep{pendall2010resilience, christopherson2010specialissue, Martin2012regional}. The notion of resilience refers to the capacity of socio-economic systems, such as sectors or regional economies, to recover from shocks and develop capabilities to navigate future challenges~\citep{fromhold2015sectoral, sutton2022regional}. As such, resilience can offer significant analytical potential in addressing one of the most compelling questions in economic geography and regional studies: Why do some regions withstand crises well while others are vulnerable to them, and why do some regional economies succeed in revitalizing themselves while others remain trapped in decline?~\citep{Hassink2010resilience, brakman2015regional, dicaro2015recessions, jun2022resilience}.

Global warming is one of the significant crises humanity faces because its impacts are far-reaching and affect every aspect of the planet. Projections indicate that Earth's temperature could rise by at least 3°C by 2100 compared to present levels \citep{peri2021economic, tollefson2020hot}. While there are diverse viewpoints regarding the nature of the changes induced by climate change, one undeniable fact is that global warming is no longer a hypothetical future scenario but an immediate and significant current crisis in our lives. The prediction that the El Niño phase will intensify and persist until 2024 suggests that it could lead to new records of extreme weather, including severe heat waves, droughts, wildfires, and flooding, raising concerns among academics, government officials, and the general public \citep{WFE2024global}.

Climate change, as an ongoing reality and looming threat, has the potential to displace populations from certain areas to others, thereby either accelerating (and shaping) or stagnating regional growth and development \citep{klein2021vast, cattaneo2016migration, wegener2013future}. At the city level, population is a crucial factor for the development and long-term sustainability of the urban economy amidst the mega-trends of low birth rates and population decline. In the short term, population is closely tied to the vitality of a city, which is associated with how active and bustling a particular urban space is at different times~\citep{ravenscroft2000vitality}. Urban vitality depends significantly on the degree of crowding due to the floating population passing through~\citep{bromley2002food}. In line with this, urban mobility, which represents the movement of populations within a city, plays significant social and economic roles, such as providing access to vital opportunities and services. Ultimately, changes in population flow and population dynamics in response to climate shocks are closely linked to regional economic resilience.

One notable point is that the impacts of climate change on urban vitality, and by extension regional resilience, show disparities across various spatial units within a city. Some areas are vulnerable to crises, while others withstand them well, maintaining our way of life and even seizing the opportunity to transform into a better state. However, in the case of global crises like global warming, there is a tendency to overlook the disparities at smaller spatial units and to emphasize a 'best practice' approach, such as carbon emission controls, that the entire world should follow ~\citep{dessai2005role, prall2023socio}. This approach fails to adequately address the varied vulnerabilities and diverse effects of climate change in a city, simply because the impact of the crisis is on a global scale. However, the impact of climate change also exhibits disparities at the intra-city scale. This raises the question: What factors result in the spatial disparity in regional resilience under climate change in urban areas?~\citep{dessai2005role, jurgilevich2021assessing, prall2023socio}.


To answer the question, we look at the economic complexity of a region as an indicator of an urban area's capability to attract the population, helping regions reinforce their resilience. Given that regional resilience refers to a region's capacity to adapt to changing economic conditions and forge a new development trajectory ~\citep{Saviotti1996, Neffke2011, Hassink2010resilience, Simmie2010resilience}, scholars contend that a region's structure of economic activities is a crucial determinant of this adaptive capacity, as it depends on the region's existing resources and capabilities ~\citep{Hidalgo2007, Boschma2015resilience, Jara2018role, Cainelli2019industrial}. This region's existing resources and capabilities can be measured by the metric from the field of economic complexity~\citep{hidalgo2021complexity, Hidalgo2009}.

In their seminal paper, \cite{Hidalgo2009} introduced the economic complexity method to explain the disparity of countries' income levels. By using world trade data, they found that a country with high economic complexity is a country that produces goods and services with high complexity. In turn, a product with high economic complexity is the product that is made in a country with high economic complexity. They solve this recursive relationship between a country and a product by using the dimensionality reduction technique, deriving two metrics of economic complexity, which are the economic complexity index of a region (ECI) and that of a product (PCI). While \cite{Hidalgo2009} uses the Economic Complexity Index (ECI) to explain future economic growth in countries, other scholars have expanded the application of these metrics to explore further implications of economic structure, such as geographical disparities in income inequality~\citep{hartmann2017linking, zhu2020export, fawaz2019spatial}, human development~\citep{lapatinas2016}, and greenhouse gas emissions~\citep{can2017impact, romero2021}. Recently, \cite{chu2023unveiling} explored the impact of ECI on economic resilience, examining the role of ECI in output growth volatility at the country level in response to trade and inflation shocks. At the inter-city level, \cite{kim2022insideout} showed that the ECI of urban regions acted as a centripetal force for cities during the COVID-19 pandemic. Although researchers have started looking at the role of ECI in the resilience of urban areas, there is still a lack of research that explores the role of ECI as a factor that attracts population flow at the intra-city level. 

Here, we contribute to the literature on economic geography, regional studies, and economic complexity by examining the effect of heat waves on the urban mobility and the role of the economic complexity of a region in mitigating the negative effect of the heat wave. For this purpose, we look at the heat wave that Seoul experienced in August 2018. In August 2018, the peak temperature of Seoul was 39.6°C, and the average daily high was 33.3°C. These temperatures were significantly higher than the 20-year averages since 2000, which were 35.01°C for the highest temperature, and 30.12°C for the average daily high. During this heat wave, people were more likely to stay inside, but this tendency was not the same within the city. Despite the heat, the loss of floating population was varying over urban regions. To see the regional disparity, we analyze various urban big data, such as the location data of entire small business stores, floating population data on subway boarding and alighting activities from August 2016 to 2021, and mobile phone usage data from July to September 2018. 

According to our main findings using the subway boarding and alighting population, temperature and the population show a negative relationship in general, and the region with higher economic complexity experienced less decrease in the floating population during the heat wave. This result suggests that the more complex region in a city sustains its floating population, leading to higher vitality and a resilient regional economy despite heat wave.

The rest of this paper is organized as follows: Section 2 provides a literature review on population dynamics under climate change and economic complexity in the context of regional resilience, Section 3 summarises the data and methods, and Section 4 presents the main results and robustness check. Finally, Section 5 concludes the ideas examined in this paper.

\section{Literature review}
\subsection{Population dynamics under climate change}
Population, as both agents of production and consumption, is closely linked to the scale and characteristics of the regional economy and development. Thus, population statistics, such as census data and the number of workers, have been used as the most representative socio-economic indicators ~\citep{foley1953urban, dobson2000landscan, cottineau2019defining,  berryman2020principles}. In this regard, population dynamics and the spatial distribution of the population — where people move from and to, where they concentrate, and where they stay — have consistently gained attention at various spatial levels ~\citep{castells2020population, newman2014modelling}. Recently, climate change has been considered one of the major drivers of population dynamics~\citep{kaczan2020impact, opitz2017climate}.

The reason climate change causes substantial displacement of populations is due to disruptions in living conditions brought about by various environmental changes ~\citep{hsiang2016climate, kim2014review}. Changes in living conditions can lead to migration to better living environments in the long-term and alter people's daily life patterns in the short term. These climate-related events are already displacing millions of people worldwide each year ~\citep{kaczan2020impact}. In 2016 alone, over 24 million people were displaced by sudden-onset climate events, with an additional unknown number displaced by slow-onset hazards such as droughts ~\citep{opitz2017climate}.
 
In the climate literature, most research about population dynamics under climate change and its moderating factors focus on \textit{`human migration'} within countries, particularly from rural to urban areas ~\citep{kaczan2020impact, castells2020population, gray2009environment}. For example, ~\cite{castells2020population} suggests that rural-urban migration in Sub-Saharan Africa is a outcome of `push' factors from rural areas, driven by deteriorating agricultural conditions worsen by climate change, rather than `pull' factors provided by urban area. In a similar vein, there are several studies that support these results \citep{bates2014markets, maurel2016climate, barrios2006climatic}. Additionally, ~\cite{gray2009environment} shows that unlike the commonly assumed out-migration from rural areas due to negative environmental changes like climate change, it can vary based on land-ownership and other factors. The primary driver of climate-induced population migration is the push factor from rural areas rather than the pull of urban centers. Consequently, this indicates that the impacts of climate change differ between urban and rural areas. In other words, urban areas are able to retain more population under climate change, implying that cities may have higher economic resilience to climate change compared to rural areas.

The impact of climate change also exhibits disparities within cities, and the forces that attract, repel, and concentrate people within specific areas can vary depending on multiple moderating factors influenced by climate change. Particularly, the impact of heat waves within urban areas can directly vary in the degree of temperature increase depending on the landscape ~\citep{ghobadi2018surveying}, and indirectly show differences in response depending on the structure of economic activities or characteristics of population groups ~\citep{lemonsu2015vulnerability}. For example, the study by \cite{lemonsu2015vulnerability} used Paris as a case study and employed an interdisplinary modeling chain that included a socio-economic model and a physically-based model of urban climate to simulate air temperature in the city during heat waves. The results indicated that the variation in heat wave risk due to densification dynamics is not limited to the impact on the urban heat island effect but also depends on population exposure to heat and varies according to population distribution. This finding suggests that the impact of extreme weather events, such as heat waves, can manifest in the patterns of the floating population, that is, population flow, within a city.

However, there are few studies that examine the impact of extreme weather induced by climate change on \textit{`population flow'} within a city — specifically, the influence on population dynamics at the intra-city level related to daily movements rather than residential \textit{`migration'}, focusing on structural characteristics of economic activities as a moderating factors of urban mobility under climate change. In this context, this study propose economic complexity, a metric for the structural characteristics of economic activities within urban areas, as a moderating factor of urban mobility under climate change.

\subsection{Economic complexity and Economic resilience}
The economic complexity method introduced by \cite{Hidalgo2009} initially aimed to explain the disparity in economic growth and development across countries using world trade data. They discovered that countries with high economic complexity produce goods and services with high complexity. In turn, products with high economic complexity are manufactured in countries with high economic complexity. They solve this recursive relationship between a country and a product by applying a dimensionality reduction technique, resulting in the economic complexity index of a region (ECI) and product complexity index (PCI) for products. These indices provide aa unified measure that reflects the availability, diversity, and sophistication of the factors or inputs present in an economy. Previous literature on the relevance of productive structures produced various indicators of technological sophistication. However, most quantitative efforts did not employ iterative or dimensionality reduction methods. Instead, they relied on indicators that averaged other measures, such as data on patents, human capital, and income.

Economic complexity metrics were originally introduced using international trade data and validated by countries' ability to predict future economic growth and development. Over time, these findings were quickly replicated, exploring geographic differences in human development ~\citep{lapatinas2016}, income and gender inequality ~\citep{barza2020eci, basile2022eci, chu2020eci, hartmann2017linking}, and sustainability ~\citep{can2017eci, dong2020emission, dordmond2021complexity, fraccascia2018green, hamwey2013mapping, mealy2022eci, neagu2019link, romero2021economic}, and greenhouse gas emissions ~\citep{can2017impact, romero2021}.

The concept of resilience can be defined as a region's ability to adapt to changing economic conditions and create a new development path from an evolutionary perspective ~\citep{Saviotti1996, Hassink2010resilience, Simmie2010resilience}. This concept is deeply embedded in the region's economic structure, performance, and overall functioning ~\citep{Doran2018us, Eriksson2016resilience, Neffke2011, Boschma2013, Kogler2013, Gao2017, Jara2018role}, implying a linkage with ECI. In line with this, \cite{chu2023unveiling} recently investigated how ECI influences economic resilience, focusing on the relationship between ECI and output growth volatility in response to trade and inflation shocks at the country level. At the inter-city level, \cite{kim2022insideout} demonstrated that urban regions with higher ECI acted as a centripetal force of population for cities during the COVID-19 pandemic. While researchers have begun to examine the role of ECI in the resilience of urban areas, there is still a gap in the research regarding ECI's impact on attracting population flow at the intra-city level.

At the city level, population is essential for the growth and long-term sustainability of the urban economy, particularly given the trends of low birth rates and population decline. In the short term, population is closely tied to a city's social and economic vitality. Urban vitality depends on how active a specific urban area is at various times and locations ~\citep{ravenscroft2000vitality}, which correlates with how crowded the area is due to the transient population ~\citep{bromley2002food}. Consequently, urban mobility, which reflects the movement of people within the city, plays crucial social and economic roles by providing access to important opportunities and services. Ultimately, variations in population flow and dynamics in response to climate shocks are closely related to regional resilience.

\section{Methodology}
\subsection{Data}
\label{section_data}
First,we utilized latitude and longitude data of store locations collected by the Korean Small Enterprise and Market Service from 2016 to 2021. This dataset encompasses information about small enterprises operating in Korea, categorized into nine primary groups: travel/leisure/entertainment, real estate, retail, accommodation, sports facilities, restaurants, other living-related services, health, and education. For each enterprise, the dataset includes the store's name, address, geolocation coordinates, and the range of amenities offered. By using this location information of small shops in Seoul, we define our special unit of analysis that capture the small business cluster of a region as shown in Section~\ref{spatialUnit}\footnote{In Korea, a small enterprise is defined by the ``Act on the Protection of and Support for Micro Enterprises'' as having fewer than ten full-time workers. Our dataset focuses on supporting these businesses and differs from others (e.g.,\cite{hidalgo2020amenity}) by excluding amenities like national parks and airports. It specifically includes amenities from small enterprises.}

\begin{figure}[!htbp]
  \centering
  \includegraphics[width=1.1\textwidth]{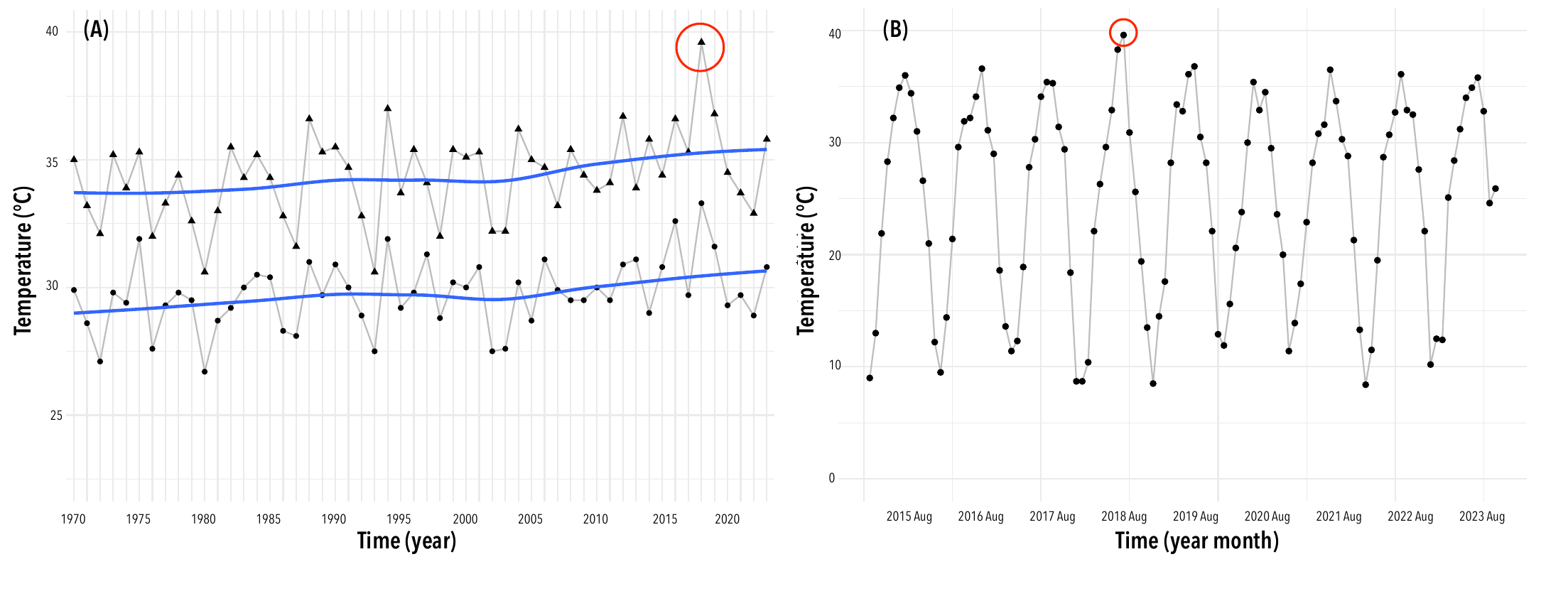}
  \caption{Temperature of Seoul: (A) Highest (triangle) and mean highest (circle) temperature in August from 1970 to 2023, and (B) monthly highest temperature from 2015 January to 2023 October. The circled points indicate the month with the highest temperature, August 2018}
  \label{fig_temp}
\end{figure}

Second, to detect the occurrence of heat stress, we use the weather data provided by the Korea Meteorological Administration. The data provide information on past weather, including daily/monthly/yearly precipitation, humidity, perceived temperature, and minimum, average, and maximum temperature. According to the weather data, from 1970 to 2022, Seoul’s mean highest and the highest temperatures in August have shown an upward trend, as depicted in Figure~\ref{fig_temp}. However, the summer of 2018 was particularly noteworthy, registering severe heatwave. In August 2018, Seoul recorded an average temperature of 28.8°C, the highest temperature of 39.6°C, and a mean daily high temperature of 33.3°C. These figures significantly exceed the 20-year average since 2000, with the August average being 26.27°C, the highest at 35.01°C, and the daily high temperature averaging 30.12°C. In this study, we see the impact of the heat stress that happened in August 2018 on the urban region, focusing on mobility and factors that effect on it.

Last, to examine the impact of high temperatures on mobility within the Seoul Metropolitan Area, we utilize the subway boarding and alighting population data from Seoul Open Data Plaza. The data includes information on the number of people getting on and off each subway line and station using transportation cards (including prepaid and disposable transportation cards) by time slot. Since cash is not acceptable for subway usage in South Korea, the data captures almost accurate population flow over time. However, this subway data is limited in that it cannot provide demographic information, such as gender and age. Additionally, this population flow data based on subway usage is limited to capturing the population flow near subway stations rather than capturing the population flow of the entire city.

To mitigate this limitation, we also use the mobility big data compiled by SK Telecom, which is the biggest network provider in South Korea. The original dataset includes a range of mobile phone activities, such as calls, texts, and internet usage, all linked to the geographical locations of users as determined by the cell towers facilitating these activities. This method enables us to track the real-time locations of subscribers. However, to adhere to privacy concerns, our data is aggregated and anonymized, using a spatial resolution of 50m x 50m unit cells to group population data. In this study, we re-aggregate the data according to the detected geographical boundaries in Section~\ref{spatialUnit} and segmented it by different gender and age groups.

Although the mobility data based on mobile phones can capture the population flow for the entire city by providing demographic information, the data is also limited in yearly comparison since the generation of mobile communication changed in December 2018. This generation change results in changing the way of calculating the population and, during a certain period, double counting the population because of the coexistence of two generations. Therefore, when we use the mobility data based on mobile phones, we only consider monthly comparisons in July, August, and September of 2018.

\begin{table}[t]
  \centering
  \caption{List of data that are used in this study}
  \scalebox{0.8}{
    \begin{tabular}{p{11.5em}p{12.835em}p{22.585em}}
    \toprule
    \toprule
    \multicolumn{1}{c}{\textbf{Data Set}} & \multicolumn{1}{c}{\textbf{Source}} & \multicolumn{1}{c}{\textbf{Explanation}} \\
    \midrule
    Small business data & The Korean Small Enterprise and Market Service & Providing information on small enterprises operating in Korea. It offers the geolocation of each shop as its longditude and latitude as well as the amenity types the shop provides. \\
    \midrule
    Weather data & The Korea Metrological Administration & Providing information on past weather, including daily/monthly/yearly precipitation, humidity, perceived temperature, and min/average/max temperature \\
    \midrule
    Subway boarding and alighting population data & Seoul Open Data Plaza & Providing information on the number of people getting on and off each subway station by time slot.  \\
    \midrule
    Floating population data based on mobile phone & SK Telecom & Providing information on the number of floating population by gender and age groups aggregated in 50m$\times$50m cell grid. Yearly comparison has been limited since the generation of mobile communication changed in December 2018.  \\
    \bottomrule
    \bottomrule
    \end{tabular}%
    }
    
  \label{tab:datalist}%
\end{table}%

Combining all the data, summarised in Table~\ref{tab:datalist}, we construct the dataset for this study. As explained above, because the mobility data based on subway usage and mobile phones have different types of limitations, we focus on only August from 2016 to 2021, when we use floating population data based on subway usage in our main analysis. When we use the data based on mobile phone usage for the robustness check, we consider the information on July, August, and September of only 2018.

In this study, the city of our interest is Seoul. Although the original source of our dataset covers the entire regions of South Korea, the mobility data is publicly available for Seoul. At the same time, Seoul, the capital city of South Korea, is a mega city, exhibiting 20\% of the total GDP of the country and covering 50\% of the total population when considering its metropolitan area. Therefore, to capture the general feature of the heat wave effect in urban areas without loss of generality, we consider only Seoul.


\subsection{Spatial unit of analysis: detecting an small business cluster}
\label{spatialUnit}
We first define the spatial unit of analysis that can capture the regional economy well. When an urban region is analyzed, administrative districts are often used as spatial units. However, considering that one of the main economic actors determining the regional economic characteristics of urban areas is small business shops, consumers or even small business owners do not align their consumption and providing services and goods with these administrative boundaries. Therefore, before moving on to the main analysis, we first define the spatial unit of analysis by looking at the location of small businesses.

We follow the methodologies of \cite{hidalgo2020amenity} and \cite{jun2022resilience} for defining the spatial unit. First, to identify amenity-dense neighborhoods, we calculate the effective number of amenities, denoted as $A_i$, for each small business location $i$, using the following equation:
\begin{equation}
 A_i = \sum^{N}_{j=1} e^{-\gamma d_{ij}}
\end{equation} where $N$ represents the total number of stores in the city, $d_{ij}$ denotes the geodesic distance between stores $i$ and $j$, and $\gamma$ is a decay parameter that reduces the influence of more distant stores. According to \cite{hidalgo2020amenity}, the value of $A_i$ decreases by half every ${\rm ln}(2)/\gamma$ kilometers. For our analysis, we use $\gamma = 7.58$, which means that the influence of an amenity halves every 91.44 meters and becomes negligible at around 804.7 meters. This distance is consistent with the median daily walking distance, as noted by \cite{Yang2012}. Therefore, $A_i$ functions as a centrality score, summarizing the number of shops within a 10-minute walking radius from store $i$~\citep{jun2022resilience}.

After calculating the effective number $A_i$ for all stores in the city, we identify the local peaks of $A_i$ on the city map. These peaks are treated as cluster centers, and we assign other stores to these clusters using an iterative greedy procedure, which incrementally expands the cluster radius, as suggested by \cite{hidalgo2020amenity}. This method resulted in the identification of 523 small business clusters within Seoul, covering an area of 605.2 $km^2$, as illustrated in Figure~\ref{fig_complexity}. The average area of these clusters is around 1.15 $km^2$, with the largest cluster radius being approximately 700 meters and the average cluster radius about 241 meters. This variation in size is due to the decreasing number of stores with increasing distance from the center and the exclusion of distant stores from the analysis. Based on this defined small business cluster as our spatial unit of analysis, we examine the variation of response upon the heatwave in August 2018.

\subsection{Economic complexity of small business clusters}

\begin{figure}[!hbp]
  \centering
  \includegraphics[width=1.1\textwidth]{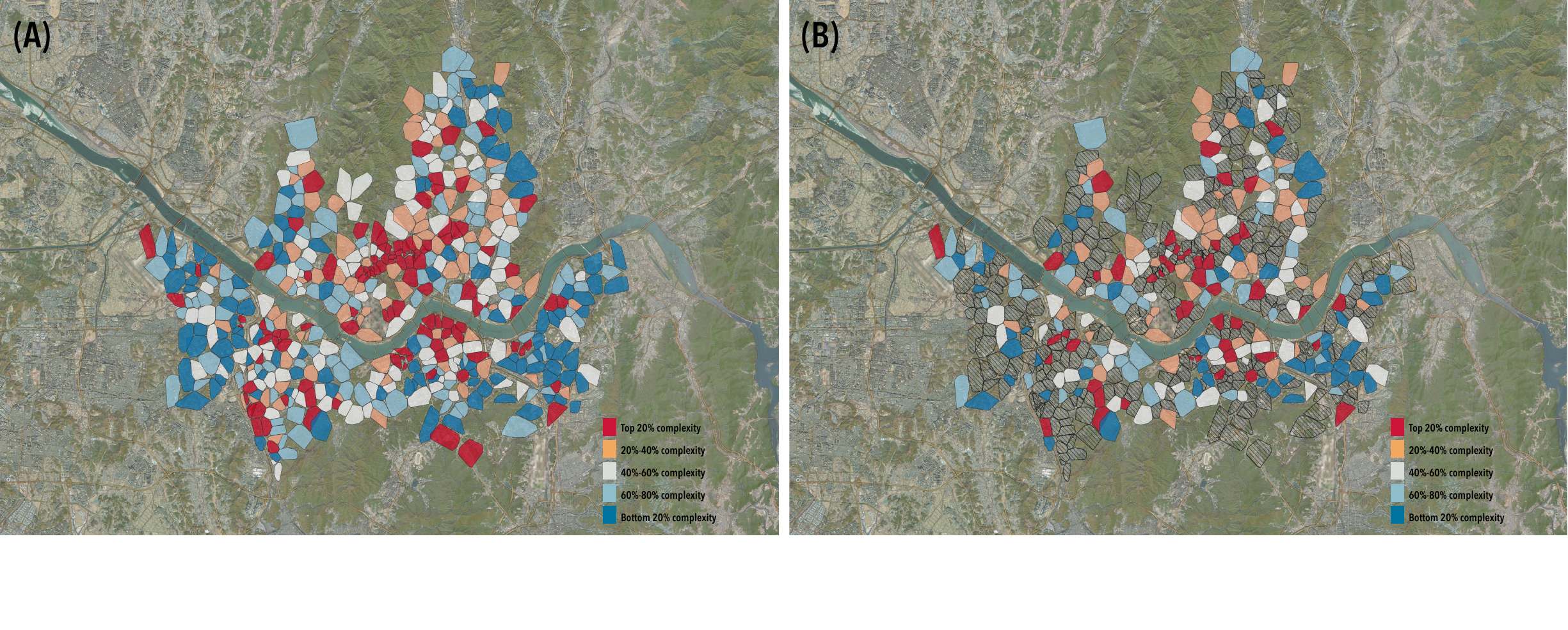}
  \caption{Complexity of small business cluster of Seoul. The spatial units of the clusters are determined by calculating the effective number of shops, denoted as, $A_i$, and the color of each cluster indicates its level of complexity (red, orange, light yellow, light blue, and navy colors represent the top 20\%, 40\%, 60\%, 80\%, and 100\% complexity, respectively. (A) displays all clusters in Seoul, while (B) specifically highlights clusters that include metro stations}
  \label{fig_complexity}
\end{figure}

This study examines the relationship between heat waves and population flow, identifying factors that mitigate the negative impact of heat on population movement and influence the economic resilience of different regions. It specifically focuses on regional vulnerability and the capacity to endure extreme high temperatures. In this regard, we propose that the economic complexity within small business clusters in various amenities plays a pivotal role in attracting people to these areas during the heat wave. Building upon the findings of \cite{kim2022insideout}, which observed that the economic complexity acts as a central force, drawing human mobility towards city centers, a trend that persisted even during the COVID-19 pandemic. However, our approach differs from that of \cite{kim2022insideout} in methodology for calculating economic complexity. While their study assessed economic complexity within administrative boundaries by looking at firms' location, we take a different approach by evaluating it based on the co-location of amenity shops, which is a foundation of the regional economy, within the boundaries of small business clusters identified in our study. Moreover, while \cite{kim2022insideout} examines the effect of economic complexity on mobility, focusing on the COVID-19 effect with considering the share of remote workers in a region, we examine the effect on mobility during the heat wave. In fact, in the era of double crises, global warming, and an extremely low birth rate, figuring out the factors that attract the population is a critical question for the development of the region.

To measure complex economic activities in an small business cluster, we calculate the economic complexity index (ECI) of each spatial unit following the method of ~\cite{Hidalgo2009} and \cite{albeaik2017improving}. Initially, we construct a bipartite network, $ M_{ci}$, consisting of cluster $c$ and industry classification of each amenity shop $i$ located in the region. Subsequently, using the method of reflections, we calculate the ECI of each region. Formally, the equation of ECI is following:
\begin{equation}
\label{eq:eci}
 \begin{split}
ECI_{c} =& 
\frac{1}{k_{c,0}} \sum_i M_{ci} k_{i, N-1} \\ 
ECI_{i} = k_{i,N} &= \frac{1}{k_{i,0}} \sum_i M_{ci} k_{c, N-1} \\ 
k_{c,0} &= \frac{1}{k_{c,0}} \sum_i M_{ci}  \\ 
k_{i,0} &= \frac{1}{k_{i,0}} \sum_i M_{ci}  \\ 
 \end{split}
\end{equation} where $ M_{ci}$ represents the small business cluster-industry bipartite network, $k_{c,0}$ and $k_{i,0}$ represent the observable levels of economic diversification of a cluster and the ubiquity of an economic activity, respectively. $ECI_{c}$ can be found as a solution of Equation~\ref{eq:eci} over a series of $N$ iterations. Although we can calculate the $ECI_{c}$ and $ECI_i$ simultaneously, our analysis focuses solely on $ECI_{c}$, which is the economic complexity index of small business clusters. The results are depicted in Figure~\ref{fig_complexity}.

Figure~\ref{fig_complexity} depicts the level of ECI of each small business cluster with the defined spatial unit of analysis. Figure~\ref{fig_complexity} (A) covers the entire area of Seoul, while (B) shows spatial units including subway station. In both figures, colors represent the level of complexity: red, orange, white, light blue, and blue color represent top 20\% of ECI, 20--40\%, 40--60\%, 60--80\%, and 80--100\%, respectively.

\subsection{Econometric model}
We construct the following empirical specification using ECI of small business clusters to examine the effect of extremely hot weather on mobility among small business clusters. Again, we cover the human mobility on each small business cluster, focusing only on August from 2016 to 2021 in our main analysis because the effect of temperature on mobility depends on the season. During the winter season, the temperature can be positively correlated with human mobility, while it can be negatively correlated with human mobility in the summer season. Also, one of our data, which is subway boarding and alighting population data, allows us to see only after 2016 to 2021 for August as well.

Equation~\ref{eq_empiric1} shows our empirical strategy using subway boarding and alighting population data, to examine the effect of heat waves on population flow in Seoul, identifying the mitigation effect of the economic complexity of a region on the flow.

\begin{equation}
\label{eq_empiric1}
  \begin{aligned}
Y_{jt} = & \beta_0 + \beta_1 Temperature_t + \beta_2 High\_Year_t + \beta_3 Rain_t + \beta_4 Covid\_period_t \\
+&\beta_5 Complexity_{jt} +\beta_6 (High\_Year_t \times Complexity_{jt}) \\
+&\beta_7 Diversity_{jt} +
\beta_8 (High\_Year_t \times Diversity_{jt}) +
\beta_{9} Total\_Shop_j   + \epsilon_{jt}
     \end{aligned}
\end{equation} where $Y_{jt}$ is aggregated floating 1,000 population of a small businesses' cluster $j$ in year $t$, $Temperature_t$ is the monthly mean highest temperature of August in year $t$, and $Highest\_Year_t$ is a binary variable indicating the year of the highest temperature, 2018 (1 for August 2018, 0 otherwise). $Rain_t$ is the monthly precipitation of August in year $t$. Considering the COVID--19 period, which covers 2020 and 2021, we add the COVID-19 dummy variable, $Covid\_period_t$. The main explanatory variable is $Complexity_{jt}$, representing the ECI value of small business cluster $j$ in year $t$, rescaled from 0 to 100. To account for the interaction between $High\_Year_t$ and $Complexity_{jt}$, we include the interaction term $High\_Year_t \times Complexity_{jt}$. Therefore, the coefficient $\beta_5$ plus $\beta_6$ shows the effect of complexity on human mobility in August 2018, when Seoul experienced the heat wave. Since one can argue that diversity of small businesses providing the urban amenity also plays a role for attracting people, we have added the diversity of a region as well. $Diversity_{jt}$ and $High\_Year_t \times Diversity_{jt}$ represent the diversity of an small business cluster $j$ in year $t$ and the interaction between the heatwave and diversity, respectively. We control for the size of the small business cluster by adding $Total\_Shop_j$, representing the total number of small businesses shops providing the urban amenity in cluster $j$. Table~\ref{tb_summary} depicts summary statistics for our variables, while Table~\ref{A_corr_metrodata} in the Appendix shows the correlation coefficients among variables. 

\begin{table}[!htbp] \centering 
\caption{Summary of descriptive statistics of the dependent and exploratory variables of the main analysis that are using metro boarding and alighting population data}
\label{tb_summary} 
\scalebox{0.75}{
\begin{tabular}{@{\extracolsep{5pt}}lccccc} 
\\[-1.8ex]\hline 
\hline \\[-1.8ex] 
Statistic & \multicolumn{1}{c}{N} & \multicolumn{1}{c}{Mean} & \multicolumn{1}{c}{St. Dev.} & \multicolumn{1}{c}{Min} & \multicolumn{1}{c}{Max} \\ 
\hline \\[-1.8ex] 
   $Y_{jt}$ (Dependent var. for time 07-24) & 1,380 & 1,143.090 & 1,003.108 & 70.530 & 6,306.167 \\ 
   $Y_{jt}$ (Dependent var. for time 07-09) & 1,380 & 189.393 & 139.570 & 14.477 & 812.874 \\ 
   $Y_{jt}$ (Dependent var. for time 09-18) & 1,380 & 563.673 & 516.080 & 35.138 & 3,565.806 \\ 
   $Y_{jt}$ (Dependent var. for time 18-20) & 1,380 & 208.508 & 187.922 & 9.435 & 1,201.105 \\ 
   $Y_{jt}$ (Dependent var. for time: 20-24) & 1,380 & 181.516 & 184.885 & 6.777 & 1,438.448 \\ 
    $Temperature_t$ & 1,380 & 31.033 & 1.554 & 29.300 & 33.300 \\ 
    $Covid\_period_t$ & 1,380 & 0.333 & 0.472 & 0 & 1 \\ 
    $Rain_t$ & 1,380 & 274.017 & 191.865 & 67.100 & 675.700 \\ 
    $Complexity_{jt}$ & 1,380 & 19.987 & 11.134 & 3.142 & 78.173  \\ 
    $Diversity_{jt}$ & 1,380 & 31.569 & 6.396 & 8 & 50 \\ 
    $Total\_Shops_{jt}$ & 1,380 & 854.263 & 462.897 & 54 & 3,127 \\ 
\hline \\[-1.8ex] 
\end{tabular} 
}
\end{table}

\section{Results}
\subsection{The effect of heat stress on population flow near subway station}

Table~\ref{tb_1_metro} depicts the results of the empirical models, Equation~\ref{eq_empiric1}. As observed in from Column (1) to (4), the mean high temperature in August negatively affects on the floating population of a region. As the coefficient of \textit{Highest Temp year} is significant and positive in Column (1), (2), and (4), it tells us that people visit small business clusters in 2018 more than other year during 2016 to 2021. At glance, it seems that it means floating population is bigger with the heat wave, but this positive sign is mainly due to the decrease of floating population during the pandemic.

The coefficient of \textit{Rain} is significant and negative for Column (2) to (4) when controlling for \textit{Complexity}. It means that rain restricts people's mobility. Next, the coefficient of \textit{COVID Period} is consistently significant and negative, meaning that the pandemic decreases the mobility of people significantly.

The explanatory variable of our interest, which is \textit{Complexity}, shows significantly positive effect on the floating population in August, meaning that people are more likely to visit place with higher complexity in general. To check whether complexity of a region plays a role in mitigating population decreasing effect during the heat wave, we have added the interaction term, \textit{High Year $\times$ Complexity}. It shows a significant and positive effect, indicating that economic complexity of a region is likely to mitigate the negative effect of heat wave on population flow. Regarding the effect of \textit{Diversity} on the population flow, interestingly, it shows significant and negative effects on the floating population in August, implying that people are less likely to visit place that provides more diverse services and products in August. Also, it doesn't show significant difference during the heat wave in 2018.

\begin{table}[!htbp] \centering 
  \caption{The effect of complexity of small business cluster on floating population (from 9 am to 24 pm) in cluster $j$ in August of year $t$}
  \label{tb_1_metro} 
\scalebox{0.8}{
\begin{tabular}{@{\extracolsep{5pt}}lcccc} 
\\[-1.8ex]\hline 
\hline \\[-1.8ex] 
 & \multicolumn{4}{c}{\textit{Dependent variable: $Y_{tj}$}} \\ 
\cline{2-5} 
\\[-1.8ex] & (1) & (2) & (3) & (4)\\ 
\hline \\[-1.8ex] 
 $Temperature_t$ & $-$65.258$^{*}$ & $-$262.608$^{***}$ & $-$238.905$^{***}$ & $-$239.309$^{***}$ \\ 
  & (36.775) & (41.153) & (42.043) & (41.853) \\ 
  & & & & \\ 
$High\_Year_t$ & 243.494$^{**}$ & 537.907$^{***}$ & 436.723 & 478.863$^{**}$ \\ 
  & (107.738) & (192.931) & (443.590) & (193.544) \\ 
  & & & & \\ 
 $Rain_t$ & $-$0.164 & $-$0.738$^{***}$ & $-$0.675$^{***}$ & $-$0.676$^{***}$ \\ 
  & (0.207) & (0.208) & (0.209) & (0.209) \\ 
  & & & & \\ 
 $Covid\_Period_t$ & $-$349.436$^{***}$ & $-$440.055$^{***}$ & $-$441.555$^{***}$ & $-$441.472$^{***}$ \\ 
  & (81.068) & (78.344) & (78.175) & (78.143) \\ 
  & & & & \\ 
 $Complexity_{jt}$ &  & 28.059$^{***}$ & 24.708$^{***}$ & 24.753$^{***}$ \\ 
  &  & (2.974) & (3.214) & (3.185) \\ 
  & & & & \\ 
 $High\_Year_t$  $\times$ $Complexity_{jt}$ &  & 13.982$^{*}$ & 13.933$^{*}$ & 13.645$^{*}$ \\ 
  &  & (7.763) & (8.214) & (7.744) \\ 
  & & & & \\ 
 $Diversity_{jt}$ &  &  & $-$12.545$^{***}$ & $-$12.343$^{***}$ \\ 
  &  &  & (4.738) & (4.331) \\ 
  & & & & \\ 
 $High\_Year_t$  $\times$ $Diversity_{jt}$ &  &  & 1.147 &  \\ 
  &  &  & (10.865) &  \\ 
  & & & & \\ 
 $Total\_Shops_j$ & 0.557$^{***}$ & 0.400$^{***}$ & 0.456$^{***}$ & 0.456$^{***}$ \\ 
  & (0.058) & (0.058) & (0.061) & (0.061) \\ 
  & & & & \\ 
 Intercept & 2,813.417$^{**}$ & 8,609.812$^{***}$ & 8,283.419$^{***}$ & 8,288.917$^{***}$ \\ 
  & (1,163.174) & (1,277.004) & (1,280.191) & (1,278.670) \\ 
  & & & & \\ 
\hline \\[-1.8ex] 
Observations & 1,380 & 1,380 & 1,380 & 1,380 \\ 
Adjusted R$^{2}$ & 0.603 & 0.635 & 0.636 & 0.637 \\ 
Residual Std. Error & 957.993 & 918.952  & 916.907  & 916.577  \\ 
 & (df = 1374) & (df = 1372) & (df = 1370) & (df = 1371) \\ 
F Statistic & 350.454$^{***}$ & 300.801$^{***}$ & 242.528$^{***}$  & 269.669$^{***}$ \\ 
 & (df = 6; 1374) & (df = 8; 1372) & (df = 10; 1370) & (df = 9; 1371) \\ 
\hline 
\hline \\[-1.8ex] 
 & \multicolumn{4}{r}{\textit{Note: Robust standard errors are reported in parentheses}} \\ 
  & \multicolumn{4}{r}{\textit{and $^{*}$p$<$0.1; $^{**}$p$<$0.05; $^{***}$p$<$0.01}} \\ 
\end{tabular} 
}
\end{table}

Since the advantage in using the subway data lies in its provision of information on the floating population by time slots, we conducted a similar analysis by time slots. The results are presented in Table~\ref{tb_2_metro}. Once again, the mean high temperature in August and the occurrence of COVID-19 pandemic decreased the floating population of a region in all time slot. Interestingly, the effects of both temperature and COVID-19 are strongest during the daytime (i.e., from 9 am to 6 pm) and smallest during morning commuting hours (i.e., 7 am to 9 am). This indicates that heat in summer has a relatively small effect on the commuting population but gives stronger negative impact on daytime floating population. Considering that the consumers of small business shops in a city are more likely to visit during working hours, the negative effect of heat on amenity shops and small business cluster based on them can be more pronounced during these hours than the effect for entire hours. 

\begin{table}[!h] \centering 
  \caption{The effect of complexity of small business cluster on floating population in cluster $j$ in August of year $t$ by time slots}
  \label{tb_2_metro} 
\scalebox{0.8}{
\begin{tabular}{@{\extracolsep{5pt}}lccccc} 
\\[-1.8ex]\hline 
\hline \\[-1.8ex] 
 & \multicolumn{5}{c}{\textit{Dependent variable: $Y_{tj}$}} \\ 
\cline{2-6} 
 \cline{2-6} 
\\[-1.8ex] & (1) 07-24 & (2) 07-09 & (3) 09-18 & (4) 18-20 & (5) 20-24\\ 
\hline \\[-1.8ex] 
$Temperature_t$ & $-$239.309$^{***}$ & $-$28.576$^{***}$ & $-$135.905$^{***}$ & $-$44.335$^{***}$ & $-$30.492$^{***}$ \\ 
  & (41.853) & (6.058) & (21.308) & (7.917) & (7.739) \\ 
  & & & & & \\ 
$High\_Year_t$ & 478.863$^{**}$ & 69.862$^{**}$ & 263.247$^{***}$ & 94.937$^{***}$ & 50.817 \\ 
  & (193.544) & (28.012) & (98.538) & (36.611) & (35.787) \\ 
  & & & & & \\ 
 $Rain_t$ & $-$0.676$^{***}$ & $-$0.082$^{***}$ & $-$0.409$^{***}$ & $-$0.120$^{***}$ & $-$0.065$^{*}$ \\ 
  & (0.209) & (0.030) & (0.106) & (0.040) & (0.039) \\ 
  & & & & & \\ 
 $Covid\_Period_t$ & $-$441.472$^{***}$ & $-$51.712$^{***}$ & $-$213.303$^{***}$ & $-$75.330$^{***}$ & $-$101.128$^{***}$ \\ 
  & (78.143) & (11.310) & (39.784) & (14.781) & (14.449) \\ 
  & & & & & \\ 
 $Complexity_{jt}$ & 24.753$^{***}$ & 2.990$^{***}$ & 14.156$^{***}$ & 4.509$^{***}$ & 3.098$^{***}$ \\ 
  & (3.185) & (0.461) & (1.622) & (0.602) & (0.589) \\ 
  & & & & & \\ 
 $High\_Year_t$ $\times$  $Complexity_{jt}$ & 13.645$^{*}$ & 0.915 & 7.954$^{**}$ & 2.403 & 2.372$^{*}$ \\ 
  & (7.744) & (1.121) & (3.942) & (1.465) & (1.432) \\ 
  & & & & & \\ 
 $Diversity_{jt}$ & $-$12.343$^{***}$ & $-$0.218 & $-$7.288$^{***}$ & $-$1.950$^{**}$ & $-$2.887$^{***}$ \\ 
  & (4.331) & (0.627) & (2.205) & (0.819) & (0.801) \\ 
  & & & & & \\ 
 $Total\_Shops_{jt}$ & 0.456$^{***}$ & 0.045$^{***}$ & 0.235$^{***}$ & 0.083$^{***}$ & 0.094$^{***}$ \\ 
  & (0.061) & (0.009) & (0.031) & (0.012) & (0.011) \\ 
  & & & & & \\ 
Intercept & 8,288.917$^{***}$ & 1,010.515$^{***}$ & 4,644.846$^{***}$ & 1,520.012$^{***}$ & 1,113.543$^{***}$ \\ 
  & (1,278.670) & (185.064) & (651.000) & (241.871) & (236.429) \\ 
  & & & & & \\ 
\hline \\[-1.8ex] 
Observations & 1,380 & 1,380 & 1,380 & 1,380 & 1,380 \\ 
Adjusted R$^{2}$ & 0.637 & 0.682 & 0.627 & 0.618 & 0.572 \\ 
Residual Std. Error (df = 1371) & 916.577 & 132.658 & 466.650 & 173.378 & 169.477 \\ 
F Statistic (df = 9; 1371) & 269.669$^{***}$ & 329.804$^{***}$ & 258.790$^{***}$ & 249.437$^{***}$ & 205.909$^{***}$ \\ 
\hline 
\hline \\[-1.8ex] 
 & \multicolumn{5}{r}{\textit{Note: Robust standard errors are reported in parentheses}} \\ 
  & \multicolumn{5}{r}{\textit{and $^{*}$p$<$0.1; $^{**}$p$<$0.05; $^{***}$p$<$0.01}} \\ 
\end{tabular} 
}
\end{table}

The variable of interest, $Complexity_{tj}$, aligns consistently with the previous results, exhibiting an increases in the floating population of a cluster, particularly during daytime. Once again, considering that ECI is calculated by examining the composition of amenity types in a spatial unit, the strongest positive effect of complexity from 9 am to 6 pm can be understood, because consumers may visit the small business shop more during daytime.

\subsection{Robustness check with mobile phone data}

One can worry that our analysis might be limited to small business clusters connected by subway lines and thus result in sampling bias, as the floating population data based on subway usage only captures those within reach of subway stations. To address this potential issue, we have extended our research to include a more comprehensive dataset. As explained in Section~\ref{section_data}, we additionally analyze the mobility information on mobile phone usage, compiled by SK Telecom, the dominant network provider in South Korea. This additional analysis allows us to examine human mobility across the entire city of Seoul.

While mobile phone data offers extensive insight into population movement, as described in Section~\ref{section_data}, it also shows limitations for annual comparisons, particularly around the 2018 heat wave. This limitation stems from the data collection method, which relies on cell towers proximal to mobile phone users. In December 2018, there was a generation shift of communication technology from the fourth to the fifth, while the third generation remained operational. This overlap and the evolving landscape of cell towers across different generations led to inconsistencies in measuring the floating population between 2018 and 2019. Therefore, in our analysis using mobile phone data for the entire of Seoul, we focus on analyzing the fluctuations in the floating population during the months of July, August, and September in 2018.
\begin{table}[!b] \centering 
  \caption{Summary of descriptive statistics of the dependent and exploratory variables of the mobile data.}
  \label{tb_summary2} 
\scalebox{0.75}{
\begin{tabular}{@{\extracolsep{5pt}}lccccc} 
\\[-1.8ex]\hline 
\hline \\[-1.8ex] 
Statistic & \multicolumn{1}{c}{N} & \multicolumn{1}{c}{Mean} & \multicolumn{1}{c}{St. Dev.} & \multicolumn{1}{c}{Min} & \multicolumn{1}{c}{Max} \\ 
\hline \\[-1.8ex] 
$Y_{jt}$ (Dependent var. for total population/1000) & 1,569 & 233.319 & 167.399 & 6.364 & 1,011.831 \\ 
$Y_{jt}$ (Dependent var. for male population/1000) & 1,569 & 136.663 & 96.818 & 3.882 & 583.344 \\ 
$Y_{jt}$ (Dependent var. for elderly population/1000) & 1,569 & 96.656 & 71.394 & 2.482 & 428.487 \\ 
$Y_{jt}$ (Dependent var. for age 60s)  & 1,569 & 34.407 & 22.810 & 1.240 & 132.419 \\ 
$High\_Temp_t$ & 1,569 & 0.333 & 0.472 & 0 & 1 \\ 
$Complexity_{jt}$ & 1,569 & 18.035 & 12.308 & 0.000 & 100.000 \\ 
$Diversity_{jt}$ & 1,569 & 30.380 & 6.902 & 3 & 47 \\ 
$Green\_Area_{jt}$ & 1,569 & 39.968 & 130.216 & 0.000 & 1,040.119 \\ 
$High\_Complexity_j$ & 1,569 & 0.199 & 0.399 & 0 & 1 \\ 
$Low\_Diversity_j$ & 1,569 & 0.065 & 0.247 & 0 & 1 \\ 
$High\_Green_j$ & 1,569 & 0.201 & 0.401 & 0 & 1 \\ 
$Near\_Metro_j$& 1,569 & 0.440 & 0.497 & 0 & 1 \\ 
\hline \\[-1.8ex] 
\end{tabular} 
}
\end{table} 
To examine the effect of the heatwave in August 2018 on the floating population and the role of economic complexity of an small business clusters concerning economic resilience by using mobile phone data, we construct the empirical model as following.
\begin{equation}
\label{eq_empiric2}
  \begin{aligned}
Y_{jt} = & \beta_0 + \beta_1 High\_Temp_t + \beta_2 High\_Complexity_j +
\beta_3 (High\_Temp_t \times High\_Complexity_j) \\
&+\beta_4 Low\_Diversity_{j} + \beta_5 (High\_Temp_t  \times High\_Complexity_{j})\\
&+\beta_6 High\_Green_{j} + 
\beta_7 (High\_Temp_t \times High\_Green_{j}) \\
&+ \beta_8 Near\_Metro_j   + \beta_9 (High\_Complexity \times Near\_Metro_j) + \epsilon_{tj}
     \end{aligned}
\end{equation} where $High\_Temp_t$ is a binary variable, 1 for Aug 2018, otherwise, 0. The distinction from Equation~\ref{eq_empiric1} lies in the representation of complexity. In yearly comparison in Table~\ref{tb_1_metro} and \ref{tb_2_metro} in the previous section, the complexity metrics are treated as variables, rescaled from 0 to 100 to capture the complexity ranking among small business clusters. However, in the monthly comparison discussed in this section, we consider the complexity value as a fixed variable over the subsequent three months. Here, $High\_Complexity_j$ is a binary variable indicating whether small business clusters $j$ have the top 20\% of complexity. Similarly, $Low\_Diversity_{j}$ is a binary variable indicating whether small business cluster $j$ ranks in the bottom 20\% in diversity. The reason why we see the effect of diversity using the bottom 20\% is that the variance of diversity among clusters is very small for the cluster with a higher value of diversity, and it is not meaningful to see the top 20\% in this case. $High\_Green_{j}$ denotes whether small business clusters $j$ encompass the top 20\% share of green areas, such as parks and mountains, over the area of the cluster. $Near\_Metro_j$ is also a binary variable, 1 when an small business cluster $j$ includes the subway station. Otherwise, 0. Table~\ref{tb_summary2} shows the summary statistics of variables.

\begin{table}[!b] \centering 
  \caption{The effect of the economic complexity of a region on floating population in cluster j in 2018 (all age and gender groups)}
  \label{tb_robust1} 
\scalebox{0.75}{
\begin{tabular}{@{\extracolsep{5pt}}lccccc} 
\\[-1.8ex]\hline 
\hline \\[-1.8ex] 
 & \multicolumn{5}{c}{\textit{Dependent variable: $Y_{jt}$ (total floating population /1000)}} \\ 
\cline{2-6} 
\\[-1.8ex] & (1) & (2) & (3) & (4) & (5)\\ 
\hline \\[-1.8ex] 
$High\_Temp_t$  & $-$6.164 & $-$5.344 & $-$5.905 & $-$6.079 &  \\ 
  & (8.223) & (9.186) & (8.509) & (9.187) &  \\ 
  & & & & & \\ 
$High\_Complexity_j$ & 134.591$^{***}$ & 147.092$^{***}$ & 135.344$^{***}$ & 130.245$^{***}$ & 130.245$^{***}$ \\ 
  & (9.739) & (14.171) & (10.191) & (9.896) & (9.892) \\ 
  & & & & & \\ 
$High\_Temp_t$ $\times$ $High\_Complexity_j$ &  & $-$4.125 &  &  &  \\ 
  &  & (20.600) &  &  &  \\ 
  & & & & & \\ 
$Low\_Diversity_{j}$ &  &  & $-$2.834 &  &  \\ 
  &  &  & (19.887) &  &  \\ 
  & & & & & \\ 
$High\_Temp_t$ $\times$ $Low\_Diversity_{j}$ &  &  & $-$3.988 &  &  \\ 
  &  &  & (33.372) &  &  \\ 
  & & & & & \\ 
$High\_Green_{j}$ &  &  &  & $-$23.374$^{*}$ & $-$23.516$^{**}$ \\ 
  &  &  &  & (11.994) & (9.852) \\ 
  & & & & & \\ 
$High\_Temp_t$ $\times$ $High\_Green_{j}$ &  &  &  & $-$0.424 &  \\ 
  &  &  &  & (20.503) &  \\ 
  & & & & & \\ 
$Near\_Metro_j$ & 88.993$^{***}$ & 94.462$^{***}$ & 88.910$^{***}$ & 87.422$^{***}$ & 87.422$^{***}$ \\ 
  & (7.831) & (8.691) & (7.843) & (7.849) & (7.846) \\ 
  & & & & & \\ 
$High\_Complexity_j$ $\times$ $Near\_Metro_j$ &  & $-$29.062 &  &  &  \\ 
  &  & (20.034) &  &  &  \\ 
  & & & & & \\ 
  Intercept & 169.473$^{***}$ & 166.693$^{***}$ & 169.544$^{***}$ & 175.721$^{***}$ & 173.694$^{***}$ \\ 
  & (6.256) & (6.632) & (6.333) & (6.918) & (6.200) \\ 
  & & & & & \\ 
\hline \\[-1.8ex] 
Observations & 1,569 & 1,569 & 1,569 & 1,569 & 1,569 \\ 
Adjusted R$^{2}$ & 0.714 & 0.714 & 0.714 & 0.715 & 0.715 \\ 
Residual Std. Error & 153.537 & 153.530 & 153.631 & 153.356  & 153.286 \\ 
&  (df = 1565) & (df = 1563) & (df = 1563) &  (df = 1563) & (df = 1565) \\ 
F Statistic & 980.539$^{***}$ & 654.110$^{***}$  & 652.903$^{***}$  & 656.183$^{***}$  & 985.038$^{***}$ \\ 
& (df = 4; 1565) & (df = 6; 1563) & (df = 6; 1563) &  (df = 6; 1563) & (df = 4; 1565) \\ 
\hline 
\hline \\[-1.8ex] 
 & \multicolumn{5}{r}{\textit{Note: Robust standard errors are reported in parentheses}} \\ 
  & \multicolumn{5}{r}{\textit{and $^{*}$p$<$0.1; $^{**}$p$<$0.05; $^{***}$p$<$0.01}} \\ 
\end{tabular} 
}
\end{table} 

The results are depicted in Table~\ref{tb_robust1}. The coefficient of $High\_Temp_t$, which indicates August of 2018, is negative but insignificant. This result doesn't show a significant difference in August, mainly because of the fact that we are comparing three months, July, August, and September, within the same year, 2018. As noticed in Figure~\ref{fig_temp}, July of 2018 also scored higher temperature than other years, although the highest temperature was in August of 2018. For example, the highest temperature on July 31 was $38.3\,^{\circ}\mathrm{C}$, compared to $39.6\,^{\circ}\mathrm{C}$ on August 1. Additionally, this monthly comparison within a year includes the seasonal difference over months. Overall, because of those reasons, $High\_Temp_t$ shows insignificant effect. However, since the negative effect of heat in August was already found through the yearly comparison of previous section, we focus on the effect of ECI more in this robustness check.

Our variable of interest, which is a potential factor that mitigate the negative effect of heat on people's mobility, is $High\_Complexity_j$. Since the complexity of a region represent the structural characteristics of a region, monthly fluctuation of this variable is not meaningful. So, the variable $High\_Complexity_j$ is not varying over the three months. Again, as $High\_Complexity_j$ shows positive and significant effect on the floating population of a small businesses' cluster, we can confirm that economic complexity of a region mitigates the negative effect of heat on the population flow. This negative effect does not show the monthly difference as the interaction term, $High\_Temp_t \times High\_Complexity_t$, is not significant, since July of 2018 scored extremely high temperature compared to other years.
\begin{table}[!htbp] \centering 
  \caption{The effect of the economic complexity of a region on a floating population over gender and ages}
  \label{tb_robust2} 
\scalebox{0.75}{
\begin{tabular}{@{\extracolsep{5pt}}lcccc} 
\\[-1.8ex]\hline 
\hline \\[-1.8ex] 
 & \multicolumn{4}{c}{\textit{Dependent variable: $Y_{jt}$ (floating population /1000)}} \\ 
\cline{2-5} 
 \\[-1.8ex]  & (1) Total(20-60) & (2) Male(20-60) & (3) Female(20-60) & (4) Elderly(60) \\ 
\hline \\[-1.8ex] 
 $High\_Complexity_j$ & 130.245$^{***}$ & 71.466$^{***}$ & 58.780$^{***}$ & 18.995$^{***}$ \\ 
  & (9.892) & (5.767) & (4.179) & (1.335) \\ 
  & & & & \\ 
$High\_Green_j$ & $-$23.516$^{**}$ & $-$12.876$^{**}$ & $-$10.640$^{**}$ & $-$1.872 \\ 
  & (9.852) & (5.744) & (4.163) & (1.330) \\ 
  & & & & \\ 
$Near\_Metro_j$  & 87.422$^{***}$ & 49.134$^{***}$ & 38.288$^{***}$ & 12.618$^{***}$ \\ 
  & (7.846) & (4.574) & (3.315) & (1.059) \\ 
  & & & & \\ 
 Intercept & 173.694$^{***}$ & 103.429$^{***}$ & 70.265$^{***}$ & 25.456$^{***}$ \\ 
  & (6.200) & (3.615) & (2.620) & (0.837) \\ 
  & & & & \\ 
\hline \\[-1.8ex] 
Observations & 1,569 & 1,569 & 1,569 & 1,569 \\ 
Adjusted R$^{2}$ & 0.715 & 0.715 & 0.709 & 0.749 \\ 
Residual Std. Error (df = 1565) & 153.286 & 89.372 & 64.766 & 20.693 \\ 
F Statistic (df = 4;1565) & 985.038$^{***}$ & 985.991$^{***}$ & 958.711$^{***}$ & 1,169.460$^{***}$ \\ 
\hline 
\hline \\[-1.8ex] 
 & \multicolumn{4}{r}{\textit{Note: Robust standard errors are reported in parentheses}} \\ 
  & \multicolumn{4}{r}{\textit{and $^{*}$p$<$0.1; $^{**}$p$<$0.05; $^{***}$p$<$0.01}} \\ 
\end{tabular}
}
\end{table} 
Column (3) of Table~\ref{tb_robust1} depicts the result with variable $Low\_Diversity_j$, confirming the results of our main analysis. Being different from our expectations, the diversity of products and services is not the factor attracting people to the region. Column (4) shows the result with variable $High\_Green_j$, indicating that people are less likely to visit green area, such as parks and mountains, under the heat wave. Initially, we anticipated a positive impact of green areas on the floating population during the summer of 2018. However, the results suggest that people are less inclined to visit small business clusters with green areas, especially during August 2018. This might be attributed to people staying indoor more during the heatwave. As shown in the last variable of Table~\ref{tb_robust1}, we have controlled the existence of metro in a region over all columns and find that people are more likely to visit a small businesses cluster that is reachable through metro.

The advantage of using mobile phone data lies in its ability to provide insights into the floating population categorized by gender and age groups. We examined the variation in effects across gender and age groups and present the results in Table~\ref{tb_robust2}. In the gender comparison, we observed no significant difference between males and females, although it's noteworthy that all coefficient for males are larger than those for females. 

In the comparison across age groups, a particularly interesting finding pertains to individuals aged more than 60 years. In comparison to other age groups, this demographic group exhibits the smallest effect of complexity on their floating population. Furthermore, individuals over 60 appear indifferent to visiting greener small business clusters during the summer of 2018. Similar to other age groups, individuals over 60 tend to visit small business clusters connected to the subway more frequently but are less inclined to visit clusters with high complexity.

\section{Conclusion}
Why do some regions withstand crises well while others do not? This question can be one of the central questions in the literature on economic resilience and regional development in the era of multi-crisis. Among crises, we investigate the impact of heat waves on urban mobility and the factors that mitigate their negative effect by looking at the heat wave that Seoul experienced in August 2018.

The main findings based on the subway boarding and alighting population data reveal that the mean high temperature in August negatively affects on the floating population of a region, and the economic complexity of a small business cluster mitigates the negative effect. The interaction effect of ECI and the abnormal heatwave in 2018 shows a significant positive impact, suggesting that high-complexity clusters exhibited lower thermal vulnerability (higher thermal resistance) in urban mobility during the heat wave in 2018. Since our data includes the pandemic periods from 2020, we can also find the negative effect of COVID-19 on the urban floating population. Interestingly, while the region's economic complexity attracts the population during the heat wave, the diversity of small business shops does not affect the region's floating population. Our robustness check based on the mobile phone data confirms our main findings.

Then, how can we explain and interpret the fact that economic complexity plays a role in attracting people to a region, different from other potential factors like diversity? Considering that the metric of economic complexity differs from others in the sense that ECI capture not only diversity but also the ubiquity and centrality of economic activity, our findings can be comprehensible. In other words, a place with a higher complexity can provide products and services that other regions cannot provide. The products and services in a region with high complexity tend to have bigger range of market boundaries. So, even during the heat wave, people visit a region with high complexity because they cannot find alternatives in other regions. This result implies the possibility that people are more concentrated in a city so that they can easily consume products and services with high complexity in the era of global warming.

While our findings provide insight the population flow within a city during the heat wave, our research is limited to showing the long-term effect of the global warming. As emphasized by ~\cite{Boschma2015resilience}, there is a need for research that integrate the short-term and the long-term perspectives to advance the understanding of regional economic resilience within an evolutionary framework. Nevertheless, due to data limitations and the ongoing nature of the global warming, our study is constrained to examining the short-term effects of a heat wave and ECI of small business clusters on urban mobility.

Additionally, in the analysis of urban mobility and regional economic resilience, it is essential take into consideration cross-space interactions. While our study utilized location data for shops to define the small business cluster and examined the impact of Economic Complexity Index within that cluster, urban mobility is influenced not only by a singular small business cluster but also by the intricate dynamics of transport networks, logistics structures, and the broader urban network shaped by population flows. Exploring the cross-space effects among these various spatial network structures is necessary for a more comprehensive understanding of spatial disparity in urban mobility under global warming trend. 

Despite our limitations, this study contributes our understanding of climate change, urban mobility, regional economic resilience and economic complexity approach and suggest the subject of future research. First of all, ~\cite{kim2022insideout, balland2020complex} showed that Economic Complexity Index works as a centripetal force of urban mobility. Our findings shed on light on the understanding of economic complexity approach by suggesting the potential significance of the Economic Complexity Index as a determinant of spatial disparity in the response to high temperatures for small business clusters, expanding upon the discussions presented in ~\cite{kim2022insideout} and ~\cite{balland2020complex}.

In addition, the impact of the global warming on urban economies has been overshadowed by the effect of other shocks such as financial crisis and COVID-19 pandemic in recent decades in the literature on regional resilience. However, global warming, broader and likely to be more enduring, is on the horizon. The challenges posed by the climate crisis are more formidable, necessitating preliminary investigation into its potential impact on regional economies and cities within the framework of regional resilience. Current climate change scenarios serve as common best practice tools for adaptation planning, involving changes in energy and transportation systems to mitigate greenhouse gas emissions ~\citep{prall2023socio, cavallaro2021climate}. However, it is essential to consider accompanying projections of future thermal vulnerabilities for the development of practical policies and planning. This study contributes to filling this research gap by analyzing practical responses to high temperatures and providing regional determinants in the context of urban thermal vulnerability for small business clusters at the intra-city level. 
 
\section*{Acknowledgement}
This project is funded by the National Research Foundation of Korea (NRF-2022R1A5A7033499 and 2022R1A2C1012895). We also acknowledge the support from Inha University. We also acknowledge Jinsu Park for data cleaning and helpful feedback.

\linespread{1.5}
\newpage

\linespread{1.5}

\newpage
\linespread{1.0}
\appendix
\section*{Appendix}

\setcounter{table}{0}
\setcounter{figure}{0} 
\renewcommand{\thetable}{A\arabic{table}}
\renewcommand{\thefigure}{A\arabic{figure}}

 \begin{table}[!htbp] \centering 
   \caption{Correlation table for main analysis using subway boarding and alighting population data} 
  \label{A_corr_metrodata} 
  \scalebox{0.8}{
\begin{tabular}{@{\extracolsep{5pt}} lccccccc} 
 \\[-1.8ex]\hline 
 \hline \\[-1.8ex] 
 & $Temp_t$ & $High\_Year_t$ & $Rain_t$ & $Covid\_Period_t$ & $Complexity_{jt}$ & $Diversity_{jt}$ & $Total\_Shops_{j}$ \\ 
 \hline \\[-1.8ex] 
 $Temp_t$ & $1$ \\ 
 $High\_Year_t$ & $0.653$ & $1$  \\ 
 $Rain_t$  & $-0.658$ & $-0.167$ & $1$ \\ 
 $Covid\_Period_t$ & $-0.698$ & $-0.316$ & $0.625$ & $1$ \\ 
 $Complexity_{jt}$ & $0.266$ & $-0.122$ & $-0.125$ & $-0.158$ & $1$  \\ 
$Diversity_{jt}$ & $0.078$ & $-0.010$ & $-0.079$ & $-0.107$ & $-0.212$ & $1$  \\ 
$Total\_Shops_{j}$ & $0.114$ & $-0.117$ & $-0.098$ & $-0.132$ & $0.356$ & $0.235$ & $1$ \\ 
 \hline \\[-1.8ex] 
\end{tabular} 
}
 \end{table}

\end{document}